\documentstyle[sprocl]{article}

\def\be{\begin{equation}}
\def\ee{\end{equation}}
\def\bea{\begin{eqnarray}}
\def\eea{\end{eqnarray}}

\def\ifmath#1{\relax\ifmmode #1\else $#1$\fi}%
\def\rd{\ifmath{{\mathrm{d}}}}

\def\lan{\langle}
\def\ran{\rangle}
\def\vs{\vskip}

\begin{document}

\title{BRANCHING PROCESSES AND KOENIGS FUNCTION}

\author{O. G. TCHIKILEV}

\address{Institute for High Energy Physics, 142281 Protvino, 
Russia\\E-mail: tchikilov@mx.ihep.su}


\maketitle\abstracts{An explicit solution of time-homogeneous pure
 birth branching processes is described.~~It gives alternative 
 extensions for the negative binomial distribution (branching processes
 with immigration) and for the Furry-Yule distribution (branching
 processes without immigration).}

\section{Introduction}

 The Furry-Yule (FY) and negative binomial (NB) distributions are
 widely used in phenomenological studies of multiplicity distributions
 at high energies.  Recently, several generalizations of the negative
 binomial distribution based on the perturbative quantum chromodynamics
 have appeared in the literature (see for example \cite{heg1,pls1}).
 The aim of this note is to present another generalization~ \cite{tch1,tch3}
 of the NB and FY distributions based on simple time-homogeneous
 branching processes,\cite{hwa1,kol1,kol2,bar1}
 more simple than the branching processes used in QCD. The NB and FY 
 distributions  occur in time-homogeneous branching processes
 with allowed transition $1\rightarrow 2$ and one can extend them allowing
 higher order transitions $1\rightarrow n$ with $n>2$. It may happen that
 these processes can be useful in the description of the hadronization stages 
 still not fully understood in the framework of the QCD.

In section~2, we describe the distribution \cite{tch1,tch3} for the pure birth 
 branching process with allowed higher-order transitions that can be used
 as an extension of the FY distribution.
 In section 3,  the distribution for the branching process with 
 immigration \cite{tch3}
 is presented. A discussion is given in  the last section.

\section{Distributions for general pure birth branching processes} 

 A branching process with continuous evolution
 parameter $t$ is determined by the  rates $\alpha_n$ for the transition 
(``splitting'') of one particle into $n$ particles with all particles 
 subsequently
 evolving independently. For a pure birth branching process
 $\alpha_0 =0$. The FY distribution occurs when only the $\alpha_2$ is not
 equal to zero.
The probability generating function $m(x,t)=\sum_{n=0}^{\infty}
 p_n(t) x^n$ satisfies the forward Kolmogorov 
 equation \cite{kol1,kol2}
 $ \frac{\partial m}{\partial t} = f(x) \frac{\partial m}{\partial x}$
 ~with
 $f(x) = \sum_{n=2}^{\infty} \alpha_n x^n - \alpha x $
 and  $\alpha = \sum \alpha_n$.
 The Taylor expansion of  this equation leads to the following
 system of equations for the probabilities $p_n$
\begin{equation}
 \frac{\rd p_1}{\rd t} = - \alpha p_1 \, \qquad , \\
\label{eq:4}
\end{equation}
 and for  $n>1$
\begin{equation}
 \frac{\rd p_n}{\rd t} = \sum_{j=1}^{n-1} j \alpha_{n-j+1} p_j - n \alpha p_n
 \qquad .
\label{eq:5}
\end{equation} 
 For the case of $N$ initial particles with
 $ p^{(N)}_n (0) = \delta _{Nn}$, the solution of this system  
  has the following form:\cite{tch3}
\begin{equation}
 p^{(N)}_n = \sum _{j=N}^{n} \pi^{(N)}_{jn} p_1^j\ ,
\label{eq:13}
\end{equation}
 with the coefficients $\pi ^{(N)}_{jn}$ obeying the recursion:
\begin{equation}
 (n-j)\pi ^{(N)}_{jn} = \sum _{l=1}^{n-j} (n-l) b_l~ \pi^{(N)}_{j(n-l)} \qquad .
\label{eq:14}
\end{equation}
 Here, $p_1 = \exp ( -\alpha t)$ and $b_l = \alpha_{l+1}/\alpha$.
 This recursion starts from $\pi^{(N)}_{NN} =1$ and the coefficient
 $\pi^{(N)}_{nn}$ can be found from the relation
 $\pi^{(N)}_{nn} = - \sum _{j=N} ^{n-1} \pi^{(N)}_{jn} $ .

 One can calculate the coefficients $\pi^{(N)}_{jn}$ using the concept of the
 Koenigs function (see \cite{kuc1,kuc2} and the references therein). 
 For the branching process starting from $N$ particles, the Koenigs function is
 defined as the limit:
\begin{equation}
 K^{(N)}(x) = K^N(x) = \lim_{n\rightarrow\infty} \frac{m^N(x,nt)}{(p_1^N)^n} =
 \sum_{j=N}^{\infty} \kappa_j^{(N)} x^j = 
 \sum_{j=N}^{\infty} \pi_{Nj}^{(N)} x^j
 \qquad . 
\label{eq:017}
\end{equation}
The recursion (\ref{eq:14}) leads to the following recurrence for the
coefficients $\kappa^{(N)}_{j}, N=1,2,...; ~j=N+1, N+2, ...$~:
\begin{equation}
 (j-N)\kappa_j^{(N)} = \sum_{l=1}^{j-N} (j-l) b_l~ \kappa^{(N)}_{(j-l)} \qquad.
\label{eq:18}
\end{equation} 
 It is convenient to denote
$ \kappa ^{(x)}_{x+n} = t_n (x)$, then $t_0 (x) =1$ and the polynomials
$t_n(x)$ are defined by the recursion:\cite{tch3}
\begin{equation}
 n t_n (x) = \sum_{l=1}^{n} (x+n-l) b_l~ t_{n-l} (x) \qquad .
\label{eq:20}
\end{equation}
 The $\kappa^{(N)}_j$ in terms of $t_n(x)$ is equal to $t_{j-N} (N)$. 

 The remarkable property of the Koenigs function is
 that it satisfies the functional Schr\"oder
equation
 $ K(m) = p_1 K(x)$.
 This can be easily obtained from the  forward Kolmogorov equation 
 $\rd m/\rd t = f(m)$. Its integration gives also  the expression
\begin{equation}
 K(x) = \exp \biggl( - \alpha \int ^{x} \frac{\rd u}{f(u)}\biggr) \qquad.
\label{eq:177}
\end{equation}

 It is convenient to introduce the function $Q(x)$, the inverse of the Koenigs
 function. Then, the inversion of the Schr\"oder equation  gives the functional
 relations
 $m(x,t)=Q(p_1 K(x))$ and
 $  m^N(x,t) = Q^N (p_1 K(x))$.
 The coefficients of the Taylor expansion for $Q^N(x)= \sum Q^{(N)}_j x^j$ 
can be found using
the following relation:
 $Q^{(N)}_j = \frac{N}{j} \kappa ^{(-j)}_{-N} $.
 One can show that this relation is equivalent to the 
 B\"urmann-Lagrange series for the inverse function (see Appendix~B in
  \cite{traub1} and the references therein).
 In terms of $t_n(x)$ it gives:
 $Q^{(N)}_j = \frac{N}{j} t_{j-N} (-j) $.
 Finally,  comparison of  equation (\ref{eq:13}) with the Taylor 
 expansion in $x$ of the $Q^N(p_1(K(x))$
leads to the following expression \cite{tch3}
 for the coefficients $\pi^{(N)}_{jn}$ in terms of $t_n(x)$:
\begin{equation}
 \pi ^{(N)}_{jn} = Q^{(N)}_j \kappa^{(j)}_n = 
\frac{N}{j} t_{j-N}(-j)t_{n-j}(j)  \qquad .
\label{eq:25}
\end{equation}

 In some cases, the functions $K(x)$ and $Q(x)$ can be found explicitly,
 for example:\\
for $b_1 =1$
\vs -3mm 
\begin{equation}
 K(x) = \frac{x}{1-x} \qquad , \qquad Q(x) = \frac{x}{1+x} \qquad ;
\label{eq:251}
\end{equation}
 for~ $b_N =1$
\vs -3mm
\begin{equation}
 K(x) = \frac{x}{(1-x^N)^{1/N}} \qquad , \qquad
 Q(x) = \frac{x}{(1+x^N)^{1/N}} \qquad;
\label{eq:252}
\end{equation}
 for $b_1+b_2 =1$
\vs -3mm
\begin{equation}
 K(x) = \frac{x}{(1-x)^{\frac{1}{1+b_2}}(1+b_2 x)^{\frac{b_2}{1+b_2}}} 
 \qquad , \qquad Q(x) = ? \qquad.
\label{eq:253}
\end{equation}
  In the general case, the expressions for $Q(x)$ and $K(x)$ are quite
 complicated.

  The solution for non-critical branching processes with non-zero
 $\alpha_0$ is an infinite series in $\exp (-\alpha ' t)$, see~ \cite{tch3}.

\section{Solution for branching processes with immigration}

 For the branching processes with immigration, there is an additional external
 source of particles appearing generally in clusters of $j$ particles with the
 differential
 rates $\beta_j$ ( $\sum \beta_j = b$ ).
 The probability generating function for the
 process starting with zero particles at $t=0$ can be written~\cite{bar1,tch2}
 as
\begin{equation}
 M(x,t) = \exp \biggl( \int_{0}^{t} g(m(x,\tau )) \rd\tau \biggr)
\label{eq:32}
\end{equation}
 with
\begin{equation}
 g(x) = \sum_{i=1}^{\infty} \beta_i x^i - b  \qquad ,
\label{eq:33}
\end{equation}
 where $m(x,\tau)$ is the solution for the underlying branching
 process without immigration.  For the underlying pure birth branching
 process. Equ.~(\ref{eq:32}) leads to the following expression:\cite{tch3}
\begin{equation}
 M(x,t) = \exp {(-bt)} \exp \biggl( \sum_{n=1}^{\infty} C_n(t) x^n \biggr)
\label{eq:34}
\end{equation}
 with
\begin{equation}
 C_n(t) = \sum_{i=1}^{n} \beta_i \sum_{j=i}^{n} \pi^{(i)}_{jn} 
 \frac{1-p_1^j (t)}
 {j\alpha} \qquad  .
\label{eq:35}
\end{equation}
 Let us denote
\begin{equation}
 \exp {( \sum_{n=1}^{\infty} C_n x^n )} = 1 + \sum_{n=1}^{\infty} V_n x^n 
 \qquad .
\label{eq:36}
\end{equation}
 Then, the final probability $P_n(t) =\exp {(-bt)} V_n(t)$. The coefficients
 $V_n(t)$ can be calculated using the recursive relation:
\begin{equation}
  n V_n = \sum _{j=1}^{n} j C_j V_{n-j} \qquad .
\label{eq:37}
\end{equation}
 This relation is known in combinatorics and is used, for example, in the
 study of combinants.\cite{comb1,comb2,comb3,comb4} The coefficients $C_n(t)$
 are known as the combinants of the multiplicity distribution $P_n(t)$.
 It is of interest to note the formal analogy between Equ.~(\ref{eq:36}) 
and the exponential representation for the $K(x)$~(\ref{eq:177}).

 The branching process with immigration leads to 
 the NB distribution  when only $\beta_1$, $\alpha_2$ and $\alpha_0$
 are not equal to zero. The resulting distribution is the Poissonian when
 only $\beta_1$ is non-zero.

\section{Discussion}

  In this note we have described explicit expressions for the  probability 
 distributions in the branching processes with the parameters
 equal to the differential
 rates $\alpha_{i+1}$ and $\beta_i$. These distributions can serve as an 
 extension for the NB and/or FY distributions when at least one of the
 coefficients with $i>1$ is non-zero. It seems natural to assume that these
 parameters have no energy dependence and therefore the energy dependence is
 concentrated in one parameter $p_1(t)=\exp (-\alpha t)$. It is of interest 
 to note that the exponential dependence on $t$ corresponds to the
 power-law dependence on the mean multiplicity $\lan n\ran$, 
since $\lan n\ran$ has linear dependence on $\exp (f'(1)t)$. 

  We have applied the  distribution with non-zero $\alpha_2$ and
 $\alpha_3$ to the e$^+$e$^-$ multiplicity data in the paper.\cite{tch1}
 It has been assumed that a fixed number $N$ of sources of particle production
 is produced at some initial stage of the interaction. These sources
 develop independently of each other according to the time-homogeneous
 branching process, producing intermediate neutral clusters. Finally,
 these clusters decay into a pair of charged or neutral hadrons with
 probabilities $\varepsilon$ or ($1 - \varepsilon$), respectively, as in the
 Goulianos model.\cite{goul} The data at c.m. energies $ \sqrt s$
 above 20~GeV
 are well described with $N=7$, with energy dependent parameter $p_1$ 
   and with the parameter $\varepsilon$ fluctuating near $0.677$. It has
 been shown that the parameter $p_1$ has power-law dependence on $\sqrt s$,
 as noted earlier.\cite{biya}
 It has been also shown that $\alpha_3$ is consistent with zero,
 but values of the ratio $\alpha_3/\alpha$ of 
 the order $\sim 0.05$ are not excluded at the present level of statistics.
  It is of interest to note that the values of $\varepsilon$ near $2/3$ are
 consistent with zero isospin of the intermediate neutral clusters and
 are different from the value $\varepsilon=0.5$ used by 
 K.~Goulianos.\cite{goul}

  It is a challenging task to derive 
  explicit expressions for the polynomials $t_n(x)$ from the recurrence
 (\ref{eq:20}). This can be done easily 
 when only one of the $b_i$ is not  equal to zero, for example,
 for $b_1=1$ the 
 $t_n (x) = \frac{x(x+1)...(x+n-1)}{n!} $.
 
 It is of interest to look for the singularities in the complex plane of the
 probability generating functions for the time-homogeneous branching
 processes (as pursued by I.~Dremin \cite{drem1} and E.~De~Wolf \cite{wolf1}).
 These singularities are connected with the singularities of the $K(x)$ and
  $Q(x)$ and, therefore, with the roots of the equation $f(x)=0$. 
 For example, for the process with $b_N=1$, the poles are:
\begin{equation}
 x_{n} = \frac{\exp (\frac{2\pi i n}{N})}{(1 - p_1^N)^{1/N}}
\label{eq:39}
\end{equation}
 ($n=1,2,...,N$). They form a regular polygon and move with $t$ to the
  circumference with unit radius.

\section*{Acknowledgments}
 I am grateful to S.~Hegyi for the opportunity to give this talk and I
 would like to thank the organizers of the Symposium for providing
 a beautiful surrounding.

\end{document}